\documentclass[12pt,aps,prb,preprint,amsmath,amssymb,bibnotes,longbibliography]{revtex4-1} 
\usepackage{graphicx,makecell} 

\bibpunct{}{}{,}{s}{}{}
\makeatletter \renewcommand\@biblabel[1]{$^{#1}$} \makeatother

\begin{document}

\title{Rotational spectra of N$_2^+$: An advanced undergraduate laboratory in atomic and molecular spectroscopy}

\author{S. B. Bayram}\email{bayramsb@miamioh.edu}
\affiliation{Miami University, Physics Department, Oxford, OH 45056
}
\author{M. V. Freamat}\email{freamamv@miamioh.edu}
\affiliation{Morrisville State College, Physics Department, Morrisville, NY 13408
}
\author{P. T. Arndt}\email{arndtpt@miamioh.edu}
\affiliation{Miami University, Physics Department, Oxford, OH 45056
}
\date{\today}

\begin{abstract}
We describe an inexpensive instructional experiment that demonstrates the rotational energy levels of diatomic nitrogen, using the emission band spectrum of molecular nitrogen ionized by various processes in a commercial AC capillary discharge tube. The simple setup and analytical procedure is introduced as part of a sequence of educational experiments employed by a course of advanced atomic and molecular spectroscopy, where the study of rotational spectra is combined with the analysis of vibrational characteristics for a multifaceted picture of the quantum states of diatomic molecules.
\end{abstract}

\pacs{0150Pa @ Laboratory experiments and apparatus} \maketitle

\section{\label{sec:introduction}Introduction}
One of the most effective pedagogical approaches to teaching upper level physics is to integrate research topics and methods into advanced laboratory courses.\cite{PCAST12,Fortenberry2007,Nagda88} For example, in the Department of Physics at Miami University, we offer four advanced laboratory courses to upper-level physics majors and graduate students.\cite{Marcum97,Marcum98,Marcum09,Bayram09,Blue10,Bayram12,Bayr12,Sand07} In one of these courses we assist students in understanding the fundamental connections between atomic and molecular spectra and the underlying structures through a series of experiments.\cite{Blue10} In this paper, we present one such experiment revealing some aspects of the quantized \textit{rotational} states of gaseous molecular nitrogen---an archetypal homonuclear diatomic molecule. The rather intuitive spectral analysis employed in this experiment provides a learning tool for the extensive concepts of quantum mechanics included in physics curricula, as well as for exciting applications such as in thermal plasmas,\cite{Moon03,Boul94} in environmental control,\cite{Mach07} or in the study of the molecular constituents of planetary and stellar atmospheres.\cite{Wyll61}  Due to its abundance in our ecosystem, the nitrogen molecule is one of the oldest topics of interest for spectrographers, with the first observations made in the second half of the 1800s.\cite{Anke70} Since then, it has become a thoroughly studied system often used for introducing the intricacies of diatomic molecular models characterized by ever surprising physics across all energy scales. Thus, while focusing on surveying its rotational structure, we note that its instructional impact and the full picture of the emission spectra of nitrogen can be fully appraised only in association with other experiments in the sequence, such as the complementary study of the \textit{vibrational} spectrum of the nitrogen diatomic molecule, described in a previous article.\cite{Bayr12}

The energy of a molecule comprises its electronic, vibrational and rotational contributions quantized into nested structures of levels with significantly different energy scales. In the experiment described in this paper, we probe the rotational states which have a level separation of the order of 0.01 eV, compared to the order of 0.1 eV for the vibrational separations, and a few eV between excited electronic states. The sample diatomic gas is nitrogen in an inexpensive commercial AC capillary discharge tube. In such a discharge, direct electron impact excitation and ionization of N$_2$ gives rise to a dauntingly complex spectrum of electronic transitions.  Since homonuclear diatomic molecules possess no permanent dipole moment, pure rotation and rotation-vibration spectra are absent.  The electronic transitions occur between a wide variety of excited molecular neutral and ion states, each with some distribution of population over the various allowed vibrational and rotational levels. The electronic structure of nitrogen has been extensively studied.\cite{Loft77,Mulliken2}

In a diatomic molecule, the overlap of atomic orbitals with similar energies spawns molecular orbitals with lower energy for constructive superposition (bonding orbital) and higher energy for destructive superposition (antibonding orbital), resulting in a characteristic layout of electronic energy levels.\cite{Herz50} The strength and length of the inter-atomic bonds depend on the particular electron configuration of these levels. Consequently, the internal energy associated with the quantized vibrational and rotational degrees of freedom of the molecule is sensitive to molecular excitations which redistribute the electron cloud around the nuclei between bonding, antibonding and unbound states.

The diatomic molecule vibrates about the equilibrium bond length corresponding to each electronic configuration, and concurrently rotates about an axis perpendicular to the bond axis through the center of mass with rotational inertia depending on the bond length. Therefore, each such electronic state contains a range of vibrational levels indexed by quantum numbers $v$ = 0, 1, 2..., and each vibrational level comprises a fine structure of rotational energy levels indexed by quantum numbers $J$ = 0,1,2,...etc. Electronic transitions therefore form bands due to changes in vibrational and rotational levels that occur during the transition.
\begin{figure*}[ht]
\centering
\includegraphics[scale=1.0]{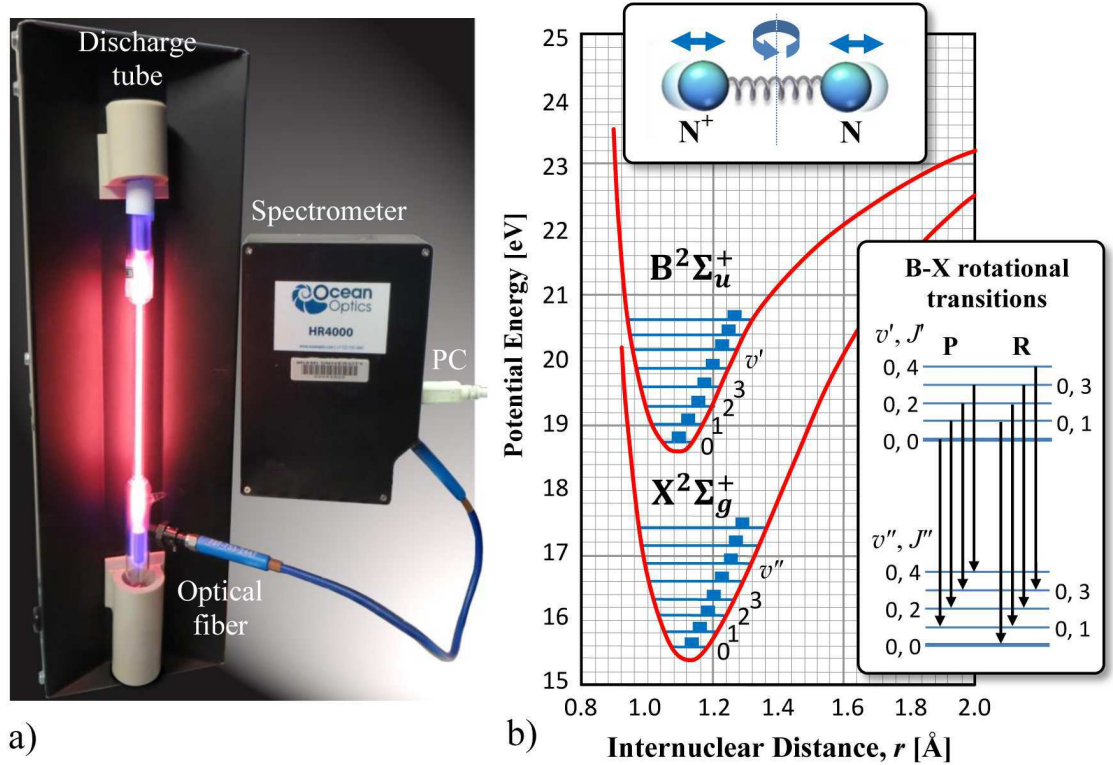}  
\caption{\label{fig:Setup} (Color online) (a) Experimental apparatus showing the fiber-spectrometer probing the emission spectrum of the nitrogen molecular ions N$_2^+$ which are created near the electrode of an AC capillary discharge tube. (b) Partial potential-energy curves of nitrogen molecular ion showing its rotational and vibrational levels in the $X^2\Sigma_g^+$ and $B^2\Sigma_u^+$ electronic states.\cite{Loft77} Due to the relation between the electron distribution in molecules and their vibrational and rotational properties, molecular electronic states contain a structure of vibrational energy levels, which in turn resolve into a fine structure of rotational levels (represented with solid rectangles on each level). Inset shows an energy level diagram for a band with P- and R-branches for the $(v'=0,J')\rightarrow (v''=0,J'')$ emission.}
\end{figure*}

Typically, the ground electronic state is labeled $X$, whereas the excited states are labeled $A$,$B$,$C$,...,$a$,$b$,$c$...,etc. In molecules, the sum of the projections of the orbital angular momenta on the line connecting the two atoms is denoted $\Lambda$. Based on $\Lambda$, the electronic states are named $\Sigma$ ($\Lambda$=0), $\Pi$ ($\Lambda$=1), and so forth. As in the case of atoms, the multiplicity of a state of a diatomic molecule is given by 2$S$+1, where $S$ is the total spin of the electrons in the molecule. The electronic state of a molecule is then labeled as $\,^{2S+1}\Lambda$. The symmetry is important when describing molecular orbitals because an electronic transition depends on whether the two orbitals involved are symmetric or antisymmetric.\cite{Herz50,Demtroder}
The positive $(+)$ or negative $(-)$ right superscript for $\Sigma$ states indicates whether the electronic wave function remains the same or changes sign by the reflection symmetry along an arbitrary plane passing through the internuclear axis. If a molecular orbital has a center of symmetry under inversion around the internuclear axis, it is designated with a subscript $g$ (``gerade'' in German, meaning even parity), and $u$ (``ungerade'' or odd parity) if it does not. The transitions are allowed for $g\leftrightarrow u$ but are forbidden for $g\leftrightarrow g$ and $u\leftrightarrow u$ since parity must change. Molecular vibrational-rotational transitions are governed by the selection rules $\Delta J = J''- J' = 0,\pm 1$ ($J'$=0$\nrightarrow$$J''=0$) where a prime ($'$) and double prime ($''$) are used to label the upper and lower electronic states respectively. In emission spectra, the transitions $\Delta J$=0 form the Q-branch, $\Delta J$=1 the P-branch, and $\Delta J$=$-$1 the R branch. A more detailed description of the selection rules governing electronic transitions and the respective spectral notation are given in Ref.[18].

In the nitrogen discharge direct electron impact ionizes the nitrogen primarily next to the electrodes, where the spectrum can be collected via optical fiber as shown in Fig.~\ref{fig:Setup}a. In our experiment, the students obtain a band of emission spectrum that results from the $X\,^2\Sigma_g^+$$\leftarrow$$B\,^2\Sigma_u^+$ transition of nitrogen molecular ion N$_2^+$ as shown in Fig.1b. One of the most prominent band systems of this transition occurs in the region 286-587 nm of N$_2^+$ and is called the \textit{first negative system}. As its name indicates, this system is observed in the negative column of a discharge through nitrogen or those containing some trace level of Argon or Helium and the bands are due to the singly positively charged molecular ion.\cite{Herz50}
Our spectrum contains the features necessary for a fairly accurate estimation of molecular  parameters.

The remainder of this paper is organized as follows. In Sect. II, we briefly introduce some theoretical background necessary to analyze the observed spectrum. Then, in Sect. III, the analysis of spectra is presented in three parts. In the first part, the students observe the spectrum and use a \textit{Fortrat diagram}\cite{Herz50} to perform a fairly accurate assignment of $J''$-values. In the second part, students are expected to analyze the spectrum, discuss its shape (such as the maximum and alternating peak intensity), and determine the rotational constants of the upper and lower electronic states. In the third part, students determine the rotational temperature of the molecule. This is followed by the conclusion in Sect. IV.

\section{\label{sec:background}Theoretical background}

The simplest quantum mechanical model of the rotating diatomic molecule envisions it as a \textit{rigid rotor}, which yields a fairly straightforward form for the rotational energy by approximating molecular vibrations and rotations as decoupled degrees of freedom. Neglecting the small centrifugal distortion caused by the stretch of the molecule (which decreases the energy slightly) the rotational energy $E_{rot}$ retains mainly its kinetic term. So, in its classical form, the energy reduces to $E_{rot}=L^2/2I$, where $L$ is the angular momentum and $I$ is the moment of inertia. According to the rigid rotor model, the molecular bond is stiff. However, in reality the bond length oscillates many times during each rotational period, so the moment of inertia $I$ can be written in terms of the average bond length $r_v$ for each allowed vibration, $I = \mu r_v^2$, with  $\mu = m_\text{N}/2 = 1.16\times10^{-26}$ kg being the reduced mass of the homonuclear diatomic molecule. Furthermore, the  quantum-mechanical nature of the molecular rotor demands that the square of the angular momentum $L^2$ take only discrete values $J(J+1)\hbar^2$, such that
\begin{equation}
\label{eq:EJ}
E_J=\frac{J\left(J+1\right)\hbar^2}{2I}.
\end{equation}
It is convenient to express the energy of the rotational level within a vibrational level $v$ in terms of wavenumber cm$^{-1}$ by a rotational term value
\begin{equation}
\label{eq:FJ}
F_v(J)=E_J/hc =B_v J\left(J+1\right)-D_v J^2(J+1)^2,
\end{equation}
where the coefficient $B_v$ is a characteristic of the vibrational state of the molecule called the \textit{rotational constant}
\begin{equation}
\label{eq:Bv}
B_v=\frac{\hbar}{4\pi c I},
\end{equation}
and $D_v$ is a centrifugal distortion constant. The magnitude of the rotational constant is a measure for the energy scale of the rotational fine structure associated with each vibrational state, and it will be different for different electronic states, even for vibrational states indexed by the same $v$. In our experiment the students are advised to ignore higher order contributions such as centrifugal distortion, except in the final comments regarding the possible corrections to the rigid-rotor model.

Equation~(\ref{eq:FJ}) yields a simple expression for the line energies in the vibrational-rotational band resulting from decays $(v',J') \rightarrow (v'',J'')$. In spectroscopic literature, transitions occurring between electronic $E_{el}$ states and vibrational $E_v$ levels of energies can also be defined in terms of a term value $T_v=( E_{el}+E_v)/hc$, with the pure vibrational lines given by $\tilde{\nu}_{v'v''} =T_{v'}-T_{v''}$. Hence, the total energy change in the transition can be written in terms of term values as $\tilde{\nu}_{v'v''J'J''}=\left(T_{v'}-T_{v''}\right) + \left(F_v'-F_v''\right)$, or
\begin{equation}
\label{eq:TransJJ}
\tilde{\nu}_{v'v''J'J''}=\tilde{\nu}_{v'v''} + B'_{v}J'\left(J'+1\right)-B''_{v}J''\left(J''+1\right).
\end{equation}
In our experiment, we observe the rotational structure for the vibrational ground states $(v',v'')=(0,0)$ and thus let $\tilde{\nu}_{v'v''}=\tilde{\nu}_{00}$. Equation~(\ref{eq:TransJJ}) can be expressed in terms of the lower quantum number $J''$ for the emission spectrum.  The P-branch means $\Delta J=1$, and so $J''=J'+1$,
\begin{equation}
\label{eq:Pbranch}
\tilde{\nu}_{P}=\tilde{\nu}_{00}-\left(B_v'+B_v''\right)J''+\left(B_v'-B_v''\right)J''^{2}.
\end{equation}
Likewise, the R-branch means $\Delta J=-1$, and so $J''=J'-1$, or
\begin{equation}
\tilde{\nu}_{R}=\tilde{\nu}_{00}+2B_v'+\left(3B_v'-B_v''\right)J''+\left(B_v'-B_v''\right)J''^{2}.
\end{equation}

\begin{figure}[!ht]
\centering
\includegraphics[scale=1.0]{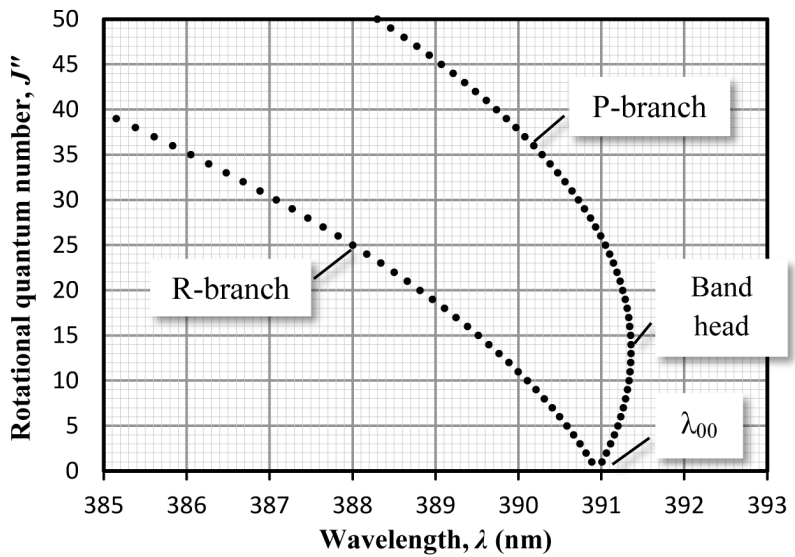} 
\caption{\label{fig:Fortrat} The Fortrat diagram of a band showing $\text{N}_2^+$ $X\,^2\Sigma_g^+(v''=0,J'')$$\leftarrow$$B\,^2\Sigma_u^+(v'=0,J')$
transitions near 391.4 nm. The circles represent transitions and the parabola guides the assignment of rotational quantum numbers $J''$ to the spectral lines in the P- and R-branches. The band head appears on the long-wavelength side of the R-branch.}
\end{figure}

The equations above (P- and R-branches) represent parabolas in $J''$ and can be plotted against the associated $J''$ on a Fortrat diagram as shown in Fig.~\ref{fig:Fortrat}. As a guide for the assignment of rotational quantum numbers $J''$ to the observed spectra, the students are provided with a Fortrat diagram generated using values for $\tilde{\nu}_{00}$, $B_v'$, and $B_v''$ from literature,\cite{Boul94} listed in Table II.
The Fortrat diagram aids in the analysis and representation of the rotational structure of molecular spectra. In the case of the first negative system, the diagram shows that the lines of the P-branch are densely distributed between the rotational quantum number 0 and 26, and the P-branch turns back, thus more transition lines form closer to the vertex of the parabola. The vertex itself corresponds to the 0-0 band head at $\lambda_{00}$=391.4 nm. The theory behind the rotational structure of diatomic molecules and Fortrat diagrams can be found elsewhere.\cite{Herz50,Bernath,Demtroder}

\section{\label{sec:analysis}Guide to the analysis of spectra}
\subsection{Analysis I: Spectral features and data}

\begin{figure*}[!ht]
\centering
\includegraphics[scale=0.9]{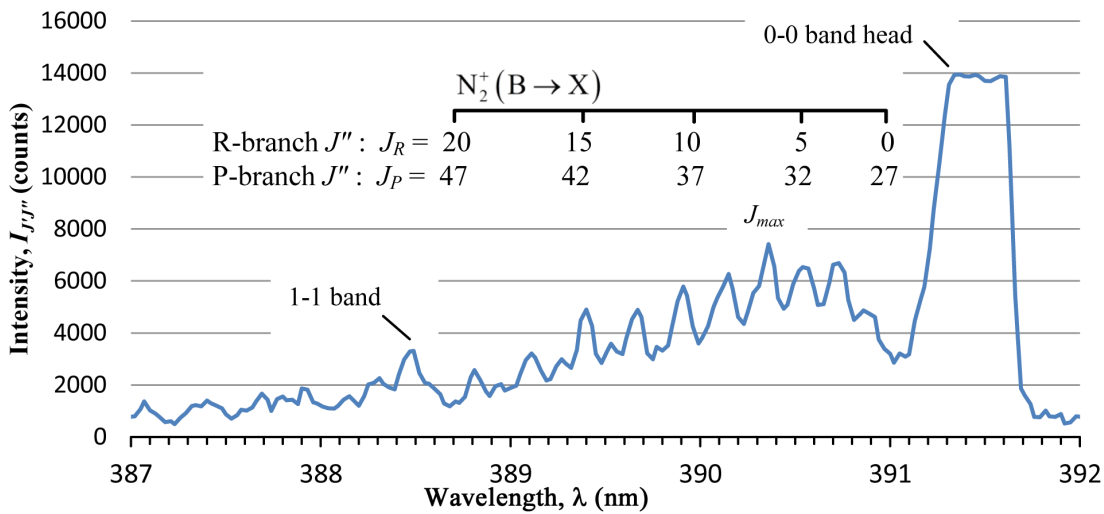} 
\caption{\label{fig:Spectrum} Measured emission band spectrum of $N_2^+$ near 391.4 nm with the fine structure of partially overlapping rotational branches quantized via rotational quantum numbers for the lower level, $J''=J_{P,R}$.}
\end{figure*}

To observe the rotational structure for the $X\,^2\Sigma_g^+$$\leftarrow$$B\,^2\Sigma_u^+$ transition, we use an Ocean Optics HR4000 spectrometer,~\cite{HR4000} calibrated using helium light,~\cite{NIST} with a resolution of 0.02 nm. Figure~\ref{fig:Spectrum} shows the observed emission spectrum of the rotational structure of the 0-0 band, where we denoted $J_P$ and $J_R$ the number $J''$ for the respective branches. Since selection rules forbid the Q-branch for transitions between $\Sigma$-states the students observed a spectrum of partially superposed P- and R-branches. It is readily observable from the spectrum that the 391.4-nm band head is unresolved due to closely-spaced by the blended $J_P$ lines, as shown on the Fortrat parabola. The 1-1 band of the first negative system is also visible at about 388.4 nm.\cite{Loft77}
The lines $J_R$ of the R-branch start close to the wavelength of the pure vibrational transition, $\lambda_{00} \approx$ 391 nm corresponding to a wavenumber $\tilde{\nu}_{00} = 25570$  cm$^{-1}$. The peaks become more separated with increasing $J$ due to the centrifugal distortion that stretches the molecule and increases the moment of inertia, thus decreasing the rotational energy. The intensity of the spectral lines $I^{em}$ in the emission spectrum is proportional to the population $N_{J'}$ of the upper electronic states and is given by
\begin{equation}
\label{eq:Intensity1}
I^{em} = g_{J'}N_{J'}A_{J'J''},
\end{equation}
where $g_{J'}$ is the statistical weight of the upper electronic states and $A_{J'J''}$  is the transition probability. Thus, because at thermal equilibrium the upper states are populated as described by Boltzmann distribution, the general tendency of rotational intensities is to form a crest about a maximum with a temperature dependent $J_{max}$. On our spectrum the maximum intensity is given by $J_R=J_{max}=7$, consistent with the value expected for room temperature according to Eq.~10. The statistical weight $g_{J'}$ includes contributions both from the $\left(2J'+1\right)$-degeneracy and the nuclear-spin parity of the upper level. The parity is symmetric for odd $J'$ with weight $\tfrac{2}{3}$, and antisymmetric for even $J'$ with weight $1/3$.\cite{Boul94, Herz50} Consequently, the intensity is expected to alternate for successive odd- and even-$J'$ lines. Note that in our case $J_P$ and $J_R$  assigned on Fig.~\ref{fig:Spectrum} represent the lower rotational levels $J''$, so one expects even-$J''$ lines to be more intense than odd-$J''$ lines. Inasmuch as the peaks result from overlapping odd with even lines from the two branches, and the statistical weight is also proportional to $2J''+1$, the even-$J_P$ lines will be the most intense.  For instance, the highest intensity line is given by an even $J_P=34$ dominating the odd $J_R=7$.

\begin{table}[!ht]
\centering
\setlength{\tabcolsep}{5pt}
\begin{tabular}{c*{5}{c}}
\hline\hline
$J_R$ & $J_P$ & $\lambda$ (nm) & $\tilde{\nu}_{J_{P,R}}$~(cm$^{-1}$)& $I^{em}$ (counts) \\
\hline
3&	30&	390.70&	25595&   6630\\
5&	32&	390.54&	25605&   6600\\
7&	34&	390.36&	25617&   7500\\
9&	36&	390.14&	25632&   6230\\
11&	38&	389.91&	25647&   5850\\
13&	40&	389.65&	25664&   4940\\
15&	42&	389.38&	25682&   4870\\
17&	44&	389.10&	25700&   3170\\
19&	46&	388.80&	25720&   2690\\
\hline\hline
\end{tabular}
\caption{The data used by the students include quantum numbers $J$ assigned using the Fortrat diagram in Fig.~\ref{fig:Fortrat}, as well as wavelengths $\lambda$ and intensities $I^{em}$ collected directly from the spectrum in Fig. 3. The standard deviation of the intensity values is about 4\%.}
\label{tab:Data}
\end{table}

After discussing the various characteristics of the spectrum, the students compute and tabulate the spectral wavenumbers ---that is, odd-$J_P$ peaks---which will suffice to estimate some rotational parameters of the molecule. Table~\ref{tab:Data} lays out values extracted from the sample spectrum shown in Fig.~3. The errors in the results are the standard deviation.

\subsection{Analysis II: Extracting rotational constants and moment of inertia}

One straightforward method is to fit one of the two data sets $(\tilde{\nu}_P,\tilde{\nu}_R )$ with a second order polynomial, compare the model coefficients with the equation for the respective branch, and build a simple system of linear equations with unknowns $B_v'$ and $B_v''$. For example, Fig.~4 shows the plot and the polynomial regression for the R-branch data sampled in Table~\ref{tab:Data}. The fit produces fairly good estimations for the rotational parameters as listed and compared with values from literature in Table~\ref{tab:Results}. Based on the rotational constants, the students calculate the excitation energy of the ion (from the free term of the polynomial fit), the moments of inertia $I'$ and $I''$ in the upper and lower electronic states (using Eq.~(\ref{eq:Bv})) and the respective average internuclear distance of the molecule. Additionally, the students can be asked to compute the band head energy by first calculating the corresponding $J_P$ from the turning point of the P-branch polynomial, and then the corresponding maximum term value (Eq.~(\ref{eq:Pbranch})).

\begin{table}[!ht]
\centering
\setlength{\tabcolsep}{5pt}
\begin{tabular}{{l}{c}{c}}
\hline\hline
Constants & Students & Literature\\
\hline
$\tilde{\nu}_{00}$~(cm$^{-1}$) & 25570$\pm$3  & 25580\cite{Boul94};~25566\cite{Loft77-1}\\
$B_v'$~~(cm$^{-1}$) & 2.014$\pm$0.110 &  2.083\cite{Boul94};~2.085\cite{Laher} \\
$B_v''$~(cm$^{-1}$) &  1.857$\pm$0.110 & 1.933\cite{Boul94};~1.932\cite{Laher}\\
$I'$~~(10$^{-46}$kg~m$^2$) & 1.39$\pm$0.08 & 1.34\cite{Boul94} \\
$I''$~~(10$^{-46}$kg~m$^2$) & 1.51$\pm$0.08 & 1.45\cite{Boul94}\\
\hline\hline
\end{tabular}
\caption{A comparison between rotational parameters estimated based on the spectrum in Fig.~\ref{fig:Spectrum} and values published in literature. The errors represent 1$\sigma$.}
\label{tab:Results}
\end{table}

\begin{figure}[!ht]
\centering
\includegraphics[scale=1.1]{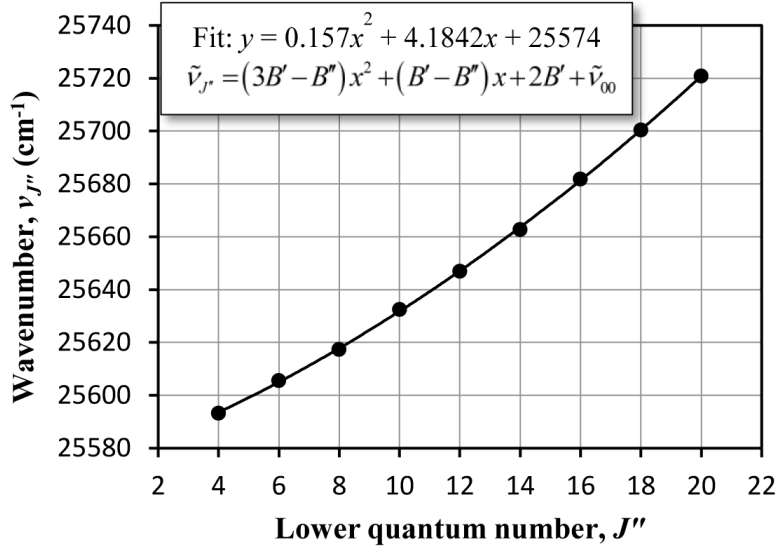} 
\caption{\label{fig:ResGraphs} The variation of the most distinct term values for the R-branch can be fit with a polynomial to obtain the rotational constants of the two states. }
\end{figure}

\subsection{Analysis III: Estimating rotational temperature}
Molecular gas temperature determination is important for the investigation of plasma processes. There are many applications of nitrogen or nitrogen containing plasmas in different types of electrical discharges. In such plasmas, the rotational distribution of nitrogen quickly achieves thermodynamic equilibrium within the gas, because nitrogen molecules exchange rotational energy faster with heavy particles than with electrons. Thus, a gas temperature can be extracted from the electronic spectra. The rovibrational band spectrum can also be used for plasmas that do not contain nitrogen as a sensitive thermometer for the gas temperature by adding a trace amount of nitrogen.\cite{Biloiu}
Electron impact excitation and ionization take place on a time scale much faster than molecular vibration or rotation.  Thus, the Boltzmann distribution of population of rotational levels in the ground state is preserved during the excitation or ionization.~\cite{Bleekrode69,Polyakova} The measured rotational temperature of the upper $B\,^2\Sigma_u^+$ state reflects the corresponding distribution in the ground state of N$_2$.

Consequently, in the third part of the analysis, the students investigate the distribution of the rotational levels to determine a rotational temperature, which is a measure of the thermal energy necessary for rotational excitations. The argument is based on Eq.~(\ref{eq:Intensity1}) for the line intensity of emission spectrum. Assuming predominantly thermal excitations, the number of molecules in the excited rotational level complies with a Boltzmann distribution weighted by the $2J+1$ degeneracy of the upper level. Because the transition also depends on the lower level, the degeneracy can be averaged between the two levels as $J'+J''+1$, such that, in a first approximation,
\begin{equation}
\label{eq:Intensity2a}
I^{em}=C\left(J'+J''+1\right)e^{-\frac{E_{J'}}{k_BT}}
\end{equation}
or
\begin{equation}
\label{eq:Intensity2}
I^{em}=C\left(J'+J''+1\right)e^{-\frac{B_vhc J'\left(J'+1\right)}{k_BT}},
\end{equation}
where $C$ is almost constant for a single band of either odd or even $J$ at constant temperature. The validity of this formula can be extrapolated to molecular excitations by direct electron impact so they hardly modify the angular momentum of the molecule. However, for excitations by collisions with heavier metastable atoms, the formula provides only a baseline model for the spectral intensities. Although some metastable Helium species may contribute to the Nitrogen ion excitation by collisions,\cite{Blue10} the model retains its pedagogical value and is sufficient to provide a good estimation for the rotational temperature. Note that the formula that maximizes the intensity is readily derivable from Eq.~(\ref{eq:Intensity2}) as
\begin{equation}
\label{eq:Jmax}
J_{max}=\sqrt{\frac{k_BT}{2B_vhc}}-\tfrac{1}{2}.
\end{equation}
Using $J_{max}=7$ and the $B_v'$ from literature, one can calculate rotational temperature of about 337 K. However, the common practice is to estimate the temperature within the context of the entire rotational band by rearranging and representing Eq.~(\ref{eq:Intensity2}) graphically as
\begin{equation}
\label{eq:TempPlot}
\text{ln}\left(\frac{I^{em}}{J'+J''+1}\right)=\text{ln}C-\frac{B_v'hc}{k_BT} J'\left(J'+1\right),
\end{equation}
where as before, $J'=J''+1$ for the R-branch and $J'=J''-1$ for the P-branch. Note that, if the left-hand term is plotted with respect to $J'(J'+1)$, the slope of the resulting linear distribution will be equal to $-B_v'hc/k_BT$, yielding the rotational temperature of the molecule. Figure~5 illustrates this method using some of the intensity values for odd $J''$ listed in Table~\ref{tab:Results} for the R-branch. The students note that, if they construct the graph within a range of low $J_R$-values, such as between 1 and 13, the slope provided by the linear fit will result in a rotational temperature of about 330$\pm$35 K. Comparing this value with the literature, the rotational temperature of the N$_2$-containing plasma in a discharge was determined in the range of 300-900 K by various methods.\cite{Janca93,Gardet2000,Cramarossa,Britun} However, counting in higher $J_R$-values will quickly decrease the slope, thus indicating that Eq.~(\ref{eq:TempPlot}) is less appropriate to model the population of higher rotational levels, especially when the excitation mechanisms cannot be confidently associated with a purely thermal distribution. Yet, this approach does render a pedagogically practical framework for the study of molecular excitations and the limitations of the rigid-rotor model.  Here the students can be encouraged to suggest explanations for the shortcomings of the model in the light of the various excitation scenarios that may be detected in the discharge tube. Related to this, they may also observe that Eq.~(\ref{eq:FJ}) is less suitable to describe rotational energies with larger $J$ when ignoring the centrifugal distortion term. This also explains why it is unreasonable to use the model with the P-branch, because on our spectrum its lines become distinct only for large $J_P$ numbers.
\begin{figure}[!ht]
\centering
\includegraphics[scale=1.2]{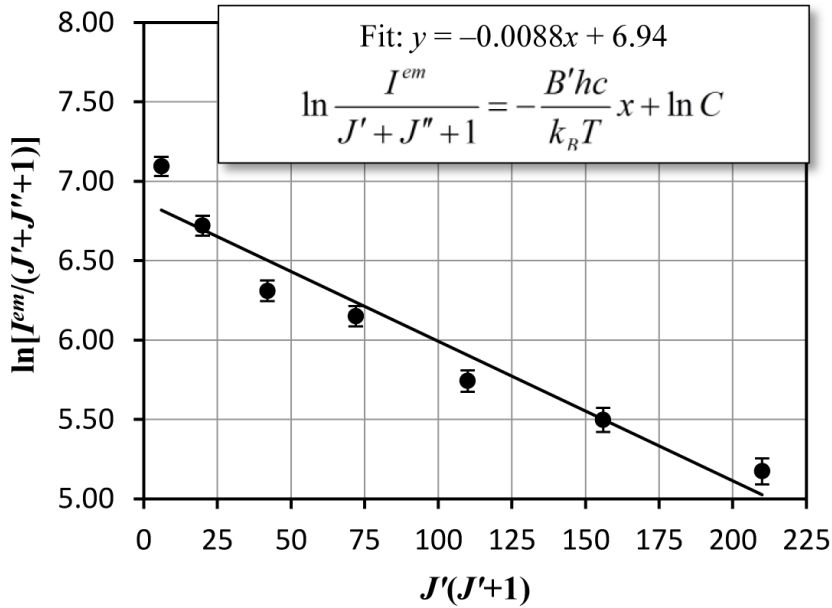}
\caption{\label{fig:ResGraphs} The rotational temperature can be estimated using a linear regression over the spectral intensity data. Considering low-$J_R$ lines, and the rotational constant $B_v'$ estimated for the $B\,^2\Sigma_u^+$ state, the slope provides a fair temperature evaluation.}
\end{figure}

\section{\label{sec:Conc}conclusions}
We described a rather straightforward and inexpensive experiment which provides an excellent demonstration of molecular spectroscopy directly in a classroom or laboratory setting. The experimental setup can shed light on some subtle aspects of the rotational structure of diatomic molecules. Whereas in a previous article we outlined an experimental procedure to observe the vibrational modes of the neutral molecule, in the present study we introduced the next logical development: a method to scrutinize the rotational modes of the ionized molecule. The data are obtained by measuring the first negative system---a particularly intense band of radiative transitions in the vicinity of 391.4 nm between rotational levels of the electronic states $B\,^2\Sigma_u^+$ and $X\,^2\Sigma_g^+$ of the nitrogen molecular ion. The ionization and excitation of nitrogen occur---likely via multiple mechanisms---in an AC capillary discharge tube, and the ensuing emission spectrum is collected using an Ocean Optics hand-held spectrometer. When combined with the rigid-rotor molecular model, the spectrum proves sufficient to extract fairly good estimations for some rotational characteristics, in particular the rotational constants and moments of inertia of the upper and lower states, as well as the rotational temperature of the molecule. In the process, the students are to make educationally formative decisions. For example, they must tabulate and plot the energies of emitted photons, assign rotational quantum numbers based on reading the provided Fortrat diagram, understand the selection rules governing the transitions, and interpret spectral features such as the band head, crest-like shape, and alternating intensities. To use the polynomial regression that yields the rotational constants and subsequently the moments of inertia, they have to choose and compare the fit with the model for one of the spectral branches. To estimate the rotational temperature, they have to first find an expected value then select a range of rotational intensities to fit linearly and extract the temperature from the slope. The analysis allows the students to observe and explain the limits of the rigid-rotor model, and the interpretation of the data allows for a more realistic perspective considering the rotational-vibrational coupling.

\begin{acknowledgments}
We gratefully acknowledge the financial support from the National Science Foundation (Grant No. NSF-PHY-1309571) and Miami University, College of Arts and Science. Also, we trace the roots of the experiment in DC discharge to the inspiring work of Doug Marcum.\cite{Blue10} Authors wish to thank Doug Marcum for insightful comments and stimulating discussions.
\end{acknowledgments}



\begin{thebibliography}{26}

\bibitem{PCAST12}President's Council of Advisors on Science and Technology, ``Engage to Excel: Producing One
Million Additional College Graduates with Degrees in Science, Technology, Engineering, and
Mathematics,'' Tech. Rep. (2012).
\bibitem{Fortenberry2007}N. L. Fortenberry, J. F. Sullivan, P. N. Jordan, and D. W. Knight, ``Engineering Education Research
Aids Instruction,'' Science {\bf 317} (5842), 1175--1176 (2007).
\bibitem{Nagda88}B. A. Nagda, S. R. Gregerman, J. Jonides, W. von Hippel, and J.S. Lerner, ``Undergraduate student-faculty research partnerships affect student retention,'' Rev. High. Educ. {\bf 22}, 55--72 (1988).
\bibitem{Marcum97}S. D. Marcum, H. Jaeger, and J. M. Yarrison-Rice, ``Advanced undergraduate laboratories at Miami University,'' presented at the Fall Meeting of the Ohio Section of the APS, Oxford, OH, 1997.
\bibitem{Marcum98}S. D. Marcum, S. G. Alexander, ``The undergraduate physics programs at Miami University:~A case study,'' presented at the AIP/AAPT Conference on Revitalization of Undergraduate Physics Programs, Alexandria, VA, 1998.
\bibitem{Marcum09}S. D. Marcum, ``Creating, implementing, and sustaining an advanced undergraduate laboratory course,'' presented at the Summer Meeting of the AAPT, Ann Arbor, MI, 2009.
\bibitem{Bayram09}S. B. Bayram, ``Vibrational spectra of the nitrogen molecules in the liquid and gas phase,'' presented at the AAPT Topical Conference on Advanced Laboratories, Ann Arbor, MI, 2009.
\bibitem{Blue10}J. Blue, S. B. Bayram, and S. D. Marcum,
``Creating, implementing, and sustaining an advanced optical spectrscopy laboratory course,'' Am.~J.~Phys. {\bf 78}, 503--509 (2010).
\bibitem{Bayram12}S. B. Bayram and M. V. Freamat, ``The Advanced Spectroscopy Laboratory Course at Miami University,''  presented at the AAPT/ALPhA Conference on Laboratory Instruction Beyond the First Year of College, Philadelphia, PA, 2012.
\bibitem{Bayr12}S. B. Bayram and M. V.  Freamat,
``Vibrational spectra of N$_2$: An advanced undergraduate laboratory in atomic and molecular spectroscopy,'' Am.~J.~Phys. {\bf 80}, 664--669 (2012)
\bibitem{Sand07}B. L. Sands, M. J. Welsh, S. Kin, R. Marhatta, J. D. Hinkle, and S. B. Bayram,
``Raman scattering spectroscopy of liquid nitrogen molecules: An advanced undergraduate physics laboratory experiment,'' Am.~J.~Phys. {\bf 75}, 488--495 (2007).
\bibitem{Moon03}S. Y. Moon, W. Choe,
``A comparative study of rotational temperatures using diatomic OH, O$_2$ and N$_2^+$ molecular spectra emitted from atmospheric plasmas,'' Spectrochim. Acta, Part B {\bf58}, 249--257 (2003).
\bibitem{Boul94} M.I. Boulos, P. Fauchais, Emil Pfender, \textsl{Thermal Plasmas: Fundamentals and Applications}, Vol 1, 1st. ed. (Springer, 1994), pp.88.
\bibitem{Mach07}Z. Machala, M. Janda, K. Hensel, I. Jedlovsky, L. Lestinska, V. Foltin, V. Martisovits, and M. Morvova,
``Emission spectroscopy of atmospheric pressure plasmas for bio-medical and environmental applications,'' J. Mol. Spectrosc. {\bf243} 194--201 (2007).
\bibitem{Wyll61}A. A. Wyller,
``Rotational Temperatures of C$_2$, CH, AlH, MgH, and SiH in Beta Pegasi,'' Astrophys J. {\bf134},  805--808 (1961).
\bibitem{Anke70}J. Anketell and J. W. Nicholls,
``The afterglow and energy transfer mechanisms of active nitrogen,'' Rep. Prog. Phys. {\bf33}, 269-306 (1970).
\bibitem{Loft77}A. Lofthus and P. H. Krupenie,
``The spectrum of molecular nitrogen,'' J.~Phys.~Chem.~Ref.~Data {\bf 6}, 113--307 (1977).
\bibitem{Mulliken2} R. S. Mulliken,
``The interpretation of Band Spectra. Parts I, IIa, IIb,'' Rev. Mod. Phys.~{\bf 2}, 60--115 (1930).
\bibitem{Herz50}G. Herzberg, \textsl{Molecular Spectra and Molecular Structure I. Spectra of Diatomic Molecules}, 2nd. ed. (Van Nostrand, Princeton, 1950).
\bibitem{Demtroder} W. Demtr\"{o}der, \textsl{Atoms, Molecules and Photons: An Introduction to Atomic-, Molecular- and Quantum Physics}, 2nd ed. (Springer, Berlin, 2006).
\bibitem{Mulliken11} R. S. Mulliken,
``Report on notation for spectra of diatomic molecules,'' Phys.~Rev.~{\bf 36}, 611--629 (1930).
\bibitem{Bernath} P. F. Bernath, \textsl{Spectra of Atoms and Molecules}, 2nd ed. (Oxford, 2005).

\bibitem{HR4000} Ocean Optics HR4000 spectrometer model with user-configured options installed: TCD1304AP (DET4-VIS) detector with VIS window with detector collection lens (L4); 5-micron slit width (SLIT-5); 2400 line holographic VIS grating (H12) with bandwidth 365-455 nm. OceanView software is used to manually configure the acquisition parameters. Including the fiber (P600-2-UV-VIS) with collimating lens (74-UV) the price for HR4000 was about $5,814.00$ in 2013.
\bibitem{NIST} NIST, physics.nist.gov/PhysRefData/Handbook/Tables.
\bibitem{Loft77-1} Reference 14, pg. 233.
\bibitem{Laher} R. R. Laher and F. R. Gilmore,
``Improved fits for the vibrational and rotational constants of many states of nitrogen and oxygen,'' J. Phys. Chem. Ref. Data~{\bf 20}, 685--712 (1991).


\bibitem{Biloiu} C. Biloiu, X. Sun, Z. Harvey and E. Scime,
``An alternative method for gas temperature determination in nitrogen plasmas: Fits of the bands of the first positive system ($B^3 \Pi_g$$\rightarrow$$A^3 \Sigma^+_u$),'' J. App. Phys.~{\bf 101}, 073303-1--073303-11 (2007).
\bibitem{Bleekrode69} R. Bleekrode and W. van Benthem,
``Spectroscopic investigations of high-current hollow-cathode discharges in flowing nitrogen at low pressures,'' J. Appl. Phys.~{\bf 40}, 5274--5280 (1969).
\bibitem{Polyakova} G. N. Polyakova, V. I. Tatus, S.S. Strel'chenko, Ya. M. Fogel and V.M. Fridman,
``Concerning the distribution of rotational energy levels of molecules excited by ion impact,'' Soviet Physics JETP~{\bf 23}, 973--978 (1966).

\bibitem{Janca93} J. Janca, L. Skricka, and A. Brablec,
``Simple and Quick Rotational Temperature Determination in N$_2$ containing Discharge Plasma,'' Plasma Chem. and Plasma Process.~{\bf 13}, 567--577 (1993).
\bibitem{Gardet2000} G. Gardet, G. Mailard, M. Courbon, F. Rogemond and M. Druetta,
``Evaluation of the rotational temperature in N$_2$ discharges using low-resolution spectroscopy,'' Meas. Sci. Technol.~{\bf 11}, 333--341 (2000).

\bibitem{Cramarossa} F. Cramarossa and G. Ferraro,
``Note: Evaluation of the rotational temperature of N$_2^+$($B^2\Sigma_u$) in an R.F. discharge at moderate pressure in nitrogen: A comparison of methods,'' J. Quant. Spectrosc. Radiat. Transfer~{\bf 14}, 159--163 (1974).

\bibitem{Britun} N. Britun, M. Gaillard, A. Ricard, Y. M. Kim, K. S. Kim, and J. G. Han,
``Determination of the vibrational, rotational and electron temperatures in N$_2$ and Ar-N$_2$ rf discharge,'' J. Phys. D: Appl. Phys.~{\bf 40}, 1022--1029 (2007).

\end{thebibliography}
\end{document}